\documentclass[10pt,twocolumn]{IEEEtran} 

\usepackage{amsmath,url}
\usepackage{epsfig,amssymb,amsbsy,verbatim,array,enumerate}
\usepackage{pstricks,psfrag,theorem,cite,paralist}

\let\intern=\iftrue

\def\figref#1{Fig.\,\ref{#1}}%
\def\E{\mathbb{E}}
\def\P{\mathbb{P}}
\def\R{\mathbb{R}}

\def\N{\mathbb{N}}

\def\ie{{\em i.e.}}
\def\eg{{\em e.g.}}

\def\var{\operatorname{var}}
\def\Beta{\beta}

\def\sir{\mathrm{SIR}}

\def\cov{\operatorname{cov}}
\def\dd{\mathrm{d}}

\def\SIR{\mathrm{SIR}}
\def\one{\mathbf{1}}

\def\eqa{\stackrel{{\rm (a)}}{=}}
\def\eqb{\stackrel{{\rm (b)}}{=}}

\newtheorem{theorem}{Theorem}

\newtheorem{corollary}{Corollary}

\newlength{\figwidth}


\begin{document}
\title{Diversity Loss due to Interference Correlation} 
\author{Martin Haenggi, University of Notre Dame
\thanks{Manuscript date \today. The partial support of the NSF (grant CNS 1016742) is gratefully acknowledged.}
}
\maketitle
\begin{abstract}
Interference in wireless systems is both temporally and spatially correlated. Yet
very little research has analyzed the effect of such correlation. Here we focus on its
impact on the diversity in Poisson networks with multi-antenna receivers. 
Most work on multi-antenna communication does not consider interference, and
if it is included, it is assumed independent across the receive antennas.
Here we show that interference correlation significantly reduces the probability of
successful reception over SIMO links. The diversity loss is quantified via the
{\em diversity polynomial}. For the two-antenna case, we provide the complete
joint SIR distribution.
\end{abstract}
\section{Introduction}
\subsection{Motivation}
Interference is a main performance-limiting factor in wireless systems.
It is spatially correlated since it stems from a single set of transmitters---even
in the presence of independent fading. It is temporally correlated since a subset
from the same given set of nodes transmits in different time slots.
While it has been long recognized that {\em correlated fading} reduces the performance
gain in multi-antenna communications, see, \eg, \cite{net:Chuah02tit},
interference correlation has been completely ignored until very recently.

In this paper, we analyze the effect of interference correlation on multi-antenna
reception in Poisson networks, where interferers form a Poisson point process (PPP),
using tools from stochastic geometry and point process theory.

\subsection{Prior work}
\subsubsection{Spatiotemporal correlation}
The first explicit results on the interference correlation in spatial networks appeared in \cite{net:Ganti09cl}.
Denoting the interference at location $x$ in time slot $m$ by $I(x,m)$,
it was shown that the temporal (Pearson's) correlation coefficient
in a Poisson network with ALOHA transmit probability $p$, unit
transmit powers, and independent and identically distributed (iid)
block fading with second moment $\E(h^2)$ is
\begin{align*}
 \rho&\triangleq\frac{\cov(I(x;m)I(x;n))}{\var I(x;m)} 
=\frac{p}{\E(h^2)}\,, \quad x\in\R^2,\, m\neq n\, ,
\end{align*}
This remarkably simple result shows
that the correlation coefficient is proportional to the transmit probability and that Rayleigh block fading
cuts the correlation to a half compared to the case of no fading. So the common randomness of the
node positions causes a significant correlation in the interference, even with severe iid fading.

\subsubsection{Local delay}
Another line of work that implicitly addresses interference correlation focuses on the local delay.
The local delay, introduced in \cite[Chap.~17]{net:Baccelli09now} and \cite{net:Baccelli10infocom} and
further analyzed in \cite{net:Haenggi12tit}, is defined
as the mean time it takes a node to successfully communicate with its nearest neighbor.
The transmission success events are correlated but they are
{\em conditionally independent} given the point process, which permits closed-form expressions
in the case of Poisson networks \cite{net:Haenggi12tit}. 

It turns out that if the transmitter density
exceeds a critical value, the correlation in the success events is strong
enough so that nearest-neighbor communication is no longer possible in finite time on
average\footnote{This does not mean that a given node cannot talk to its nearest neighbor in finite
time; it means that the number of slots until success has a heavy tail, such that the mean diverges.}.
So the local delay is not only a basic metric that quantifies the performance of a network, it is
also a sensitive indicator of correlation.

\section{System Model}
We consider a {\em Poisson network}, where the interferers, all equipped with one antenna, form a stationary
Poisson point process (PPP) $\Phi\subset\R^2$ of intensity $\lambda$.
The receiver under consideration is assumed
to be located at the origin $o$ and equipped with $n\geq 1$ antennas,
and a desired transmitter is added at distance $r$ from the origin.
All channels are subject to iid Rayleigh fading.
The SIR at antenna $k$ of the receiver is
\[ \SIR_k=\frac{h_k r^{-\alpha}}{\sum_{x\in\Phi} h_{x,k} \|x\|^{-\alpha}} \,,\quad k\in[n]\,, \]
for independent exponential $h_k, h_{x,k}$ and a path loss exponent $\alpha>2$ (otherwise the
interference would be infinite a.s. \cite{net:Haenggi08now}). $[n]$ denotes the
set $\{1,2,\ldots, n\}$.

Our main concern are the probabilities of events of the type $S_k\triangleq\{\SIR_k>\theta\}$ and unions and
intersections thereof.

 For $n=1$, it is well known that \cite{net:Haenggi08now}
 \begin{equation}
  P_1(\theta)\triangleq \P(S_1)=\exp(-\Delta \theta^\delta) \,,
  \label{ps_siso}
 \end{equation}
where $\delta\triangleq 2/\alpha$ and
 $\Delta\triangleq \lambda\pi r^2\Gamma(1+\delta)\Gamma(1-\delta)$.

\section{Diversity in SIMO System}
 Despite the independent fading, the interference at each antenna is
 correlated due to the common interferer locations,
 hence the events $S_k$ and $S_j$ are not independent. 
We focus first on the probability of their joint occurrence
\[ P_n(\theta)\triangleq\P\bigg(\bigcap_{k\in[n]} S_k\bigg) \,.\]

 \begin{figure}
 \centerline{\epsfig{file=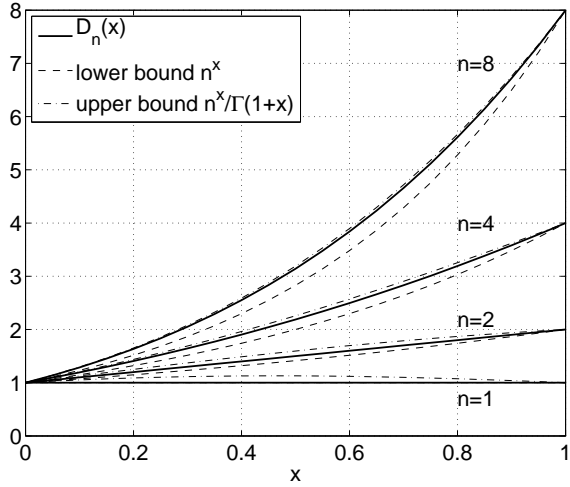,width=\figwidth}}
 \caption{Diversity polynomial $D_n(x)$, lower bound $n^x$, and upper bound $n^x/\Gamma(1+x)$
 for $n=1,2,4,8$.}
 \label{fig:dn_fig}
 \end{figure}

 \subsection{Main result}
\begin{theorem}
\label{thm:simo_prob}
 The probability that the SIR at all antennas exceeds $\theta$
 is
 \begin{equation}
  P_n(\theta)=
   \exp(-\Delta\theta^\delta D_n(\delta))\,, 
  \label{simo_joint}
\end{equation}
 where $D_n$ is the polynomial of order $n-1$ given by
 \[ D_n(x)=
 \frac{\Gamma(n+x)}{\Gamma(n)\Gamma(1+x)}=\frac{1}{x\Beta(n,x)}\,. \]
  $\Beta(x,y)\triangleq\Gamma(x)\Gamma(y)/\Gamma(x+y)$ is the Beta function.
 \end{theorem}
 \begin{IEEEproof}
 Let $\theta_r=\theta r^\alpha$. Then the SIR condition for a single antenna is 
 \[ \frac{hr^{-\alpha}}I>\theta \quad\Longleftrightarrow \quad h>\theta_r I \,,\]
 and we have
 \[ P_n(\theta)= \P(h_1>\theta_r I_1,\ldots, h_n>\theta_r I_n)\,,\]
 where $h_i$ are the iid fading coefficients to each antenna, and
 $I_k=\sum_{x\in\Phi} h_{x,k}\|x\|^{-\alpha}$ is the interference
 at each antenna,
  correlated through the common randomness $\Phi$.
  We obtain
 \begin{align*}
  P_n(\theta)& =\E(e^{-\theta_r I_1}\cdots e^{-\theta_r I_n})\\
  &=\E\prod_{k=1}^n e^{-\theta_r I_k} \\
   &=\E\prod_{k=1}^n \prod_{x\in\Phi} e^{-\theta_r h_{x,k}\|x\|^{-\alpha}} \\
   &\eqa\E\prod_{x\in\Phi} \E_h \prod_{k=1}^n e^{-\theta_r h_{x,k}\|x\|^{-\alpha}} \\
  &=\E\prod_{x\in\Phi} \left(\frac{1}{1+\theta_r\|x\|^{-\alpha}}\right)^n \\
  &\eqb \exp\left(-\lambda \int_{\R^2}\left(1-\left(\frac{\|x\|^\alpha}{\|x\|^\alpha+\theta_r}\right)^n\right)\dd x\right)\,.
 \end{align*}
 (a) follows from the independence of the fading random variables $h_{x,i}$, and (b) follows from the
 probability generating functional of the PPP.
The last step is the calculation of the integral, which yields the result.
 \end{IEEEproof}
 
\subsection{The diversity polynomial}
We term the polynomial $D$ the {\em diversity polynomial}.
 The first four are
 \begin{align*}
 D_1(x)&=1 \\
 D_2(x)&=1+x\\
 D_3(x)&=\tfrac12 (x+1)(x+2) \\
 D_4(x)&=\tfrac16 (x+1)(x+2)(x+3) \,,
 \end{align*}
 and a general expression is
 \[ D_n(x)=\frac1{\Gamma(n)}\prod_{i=1}^{n-1} (i+x)=\prod_{i=1}^{n-1} \left(1+\frac x i\right)\,. \]
  For all $x\in(0,1)$, since $\tfrac{\Gamma(n+x)}{\Gamma(n)}\lesssim n^x$,
  \begin{equation}
  n^x< D_n(x)\lesssim \frac{n^x}{\Gamma(1+x)} \,.
  \label{dn_bounds}
  \end{equation}
 ``$\lesssim$" indicates an upper bound with asymptotic equality, here as $n\to\infty$.
  The diversity polynomials for $n=1,2,4,8$ are shown in \figref{fig:dn_fig}, together with these lower
  and upper bounds.

 The polynomial may also be defined by its $n-1$ roots
 \[ D_n(x)=0 \quad\forall x\in -[n-1] \]
 and fixing either $D_n(0)=1$ or $D_n(1)=n$.
 
\begin{figure}
 \centerline{\epsfig{file=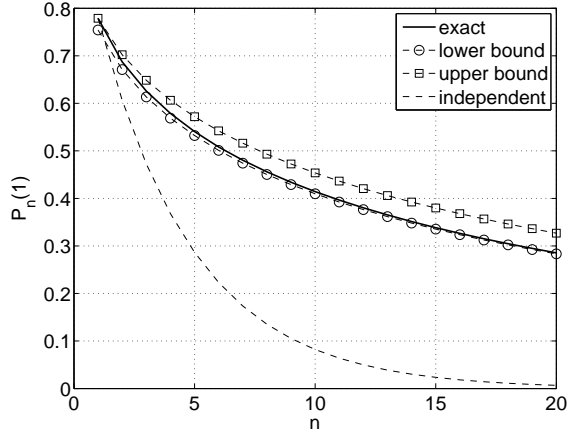,width=\figwidth}}
 \caption{$P_n(1)$ as a function of $n$ for $\Delta=1/4$ and $\delta=1/2$, together with bounds \eqref{pn_bounds},
 and the same probability under the assumption of independent interference.}
 \label{fig:pn_fig}
 \end{figure}

Since all the coefficients are positive, $D_n(x)$ is convex for $x\geq 0$ and thus
bounded by
\[ D_n(x)\leq 1+(n-1)x \,,\quad n\in\N, \,0\leq x\leq 1 \,.\]
 The derivative is asymptotically
 \begin{equation}
   D_n'(x) \triangleq \frac{\dd D_n(x)}{\dd x}=\Theta(n^x \log n)\,,\quad 0\leq x\leq 1, \,n\to\infty\,.
   \label{q_asympt}
 \end{equation}
 For $x=0$ and $x=1$, the result is exact, \ie, $D'_n(x)\sim n^x\log n$, $x\in\{0,1\}$.
From the bounds in \eqref{dn_bounds} it follows that
 \begin{equation}
  \exp(-\Delta\theta^\delta n^\delta) > P_n(\theta) > \exp\left(-\Delta\theta^\delta \frac{n^\delta}{\Gamma(1+\delta)}\right) \,. 
  \label{pn_bounds}
 \end{equation}
 $P_n(1)$ and the bounds are shown in \figref{fig:pn_fig}.

 \subsection{Diversity loss}
 If the interference was independent across the antennas, we would have 
 \[ \tilde P_n(\theta)=\exp(-\Delta\theta^\delta n)\,. \]
 
 Due to the dependence, $D_n(\delta)<n$ for all $\delta<1$ and only $D_n(1)=n$, but $\delta=1$ corresponds
 to $\alpha=2$, which would imply $\Delta=\infty$ and $P_1(\theta)=0$.
 The dependence increases as $\delta\downarrow 0$ (with growing $\alpha$). For $\delta=0$, 
 $P_n(\theta)=P_1(\theta)$, $\forall n\in\N$ (complete correlation).

 \begin{corollary}
 \label{cor:div_loss}
 The diversity loss, defined as $L(n)\triangleq\log\tilde P_n/\log P_n$, 
 is
 \[ L(n)=n\delta \beta(n,\delta)=\frac{\Gamma(n+1)\Gamma(\delta+1)}{\Gamma(n+\delta)} \,.\]
 As $n\to\infty$, $L(n)\to\infty$.
 \end{corollary}
 \begin{IEEEproof}
 From Thm.~\ref{thm:simo_prob} we obtain
 $\log \tilde P_n/\log P_n=n/D_n(\delta)$.
 For the limit, we need to show that
 \[  \lim_{n\to\infty} \frac{D_n(\delta)}{n}=0 \quad \text{ for } 0\leq \delta<1 \,. \]
This holds since
 \[ \frac{\Gamma(n+\delta)}{\Gamma(n+1)} \sim n^{\delta-1} \,,\quad n\to\infty\,,\]
 and $\delta<1$.
 \end{IEEEproof}
  The fact that $D_n(\delta)/n\to 0$ is also apparent from the asymptotic behavior of the derivative
 \eqref{q_asympt}.

 Next we determine the conditional probability that $S_{k+1}$ holds given that
$S_1,\ldots, S_k$ hold.
 
 \begin{corollary}
 \[ \P\left(S_{k+1}\mid S_1\cap \ldots \cap S_k\right)=\exp(-\Delta\theta^\delta
 D_k(\delta)\delta/k) \,,\]
 and
 \begin{equation}
  \lim_{k\to\infty} \P\left(S_{k+1}\mid S_1\cap \ldots \cap S_k\right)=1 \,.
  \label{q_limit}
 \end{equation}
 \end{corollary}
 \begin{IEEEproof}
 The conditional probability is $P_{k+1}/P_k$, which, using the recursion
 $D_{n+1}(x)=D_n(x)(1+x/n)$, yields the result.
 The limit \eqref{q_limit} follows from the proof of Cor.~\ref{cor:div_loss}.
 \end{IEEEproof}
So the correlation is strong enough that, assuming $n=\infty$, for each $\epsilon>0$, there is an $m$ such that
for any $k>m$, $S_k$ occurs with probability exceeding $1-\epsilon$ if $S_1,\ldots, S_m$ hold.
 
 \subsection{Correlation coefficients}
\begin{figure}
 \centerline{\epsfig{file=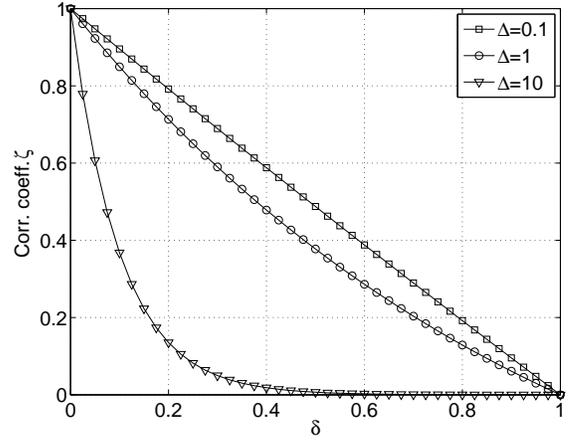,width=\figwidth}}
 \caption{Correlation coefficient $\zeta$ per \eqref{corr_coeff} of the indicators of the events $S_k$ as a function
 of $\delta$ for $\theta=1$ and $\Delta=0.1,\,1,\,10$. For small $\Delta$ or $\theta$,
 $\zeta\approx 1-\delta$.}
 \label{fig:corr_coeff}
 \end{figure}

 Let $A_k=\one\{S_k\}$ be the indicator that $S_k$ occurs. Pearson's correlation coefficient
 between $A_i$ and $A_j$, $i\neq j$, is
 \begin{equation}
  \zeta(A_i,A_j)=\frac{e^{-\Delta\theta^\delta\delta}\left(1-e^{-\Delta\theta^\delta(1-\delta)}\right)}
 {1-e^{-\Delta\theta^\delta}}\,,\quad i\neq j \,.
 \label{corr_coeff}
 \end{equation}
 The correlation coefficients for different parameters are shown in \figref{fig:corr_coeff}.
 It is easily seen that $\zeta(A_i,A_j)=1$ (full correlation) for $\delta=0$, while $ \zeta(A_i,A_j)=0$
 for $\delta=1$. So a larger path loss exponent $\alpha=2/\delta$ results in higher correlation.
 This can be explained as follows: For large $\alpha$, the interference is dominated by a few
 nearby interferers, and if one of them is close enough to cause an outage at one antenna,
 it is likely to do so also at another. Conversely, as $\alpha\downarrow 2$, the interference
 is dominated by the many far interferers, each one with an independently fading channel
 to each antenna, which decorrelates the events.
  
 In the high-reliability regime, where $\Delta$ or $\theta$ is small, the correlation is the largest;
 it is upper bounded by and approaches $1-\delta$ as $\Delta\to 0$ or $\theta\to 0$.
 
\subsection{Effect on selection combining}
\begin{figure}
 \centerline{\epsfig{file=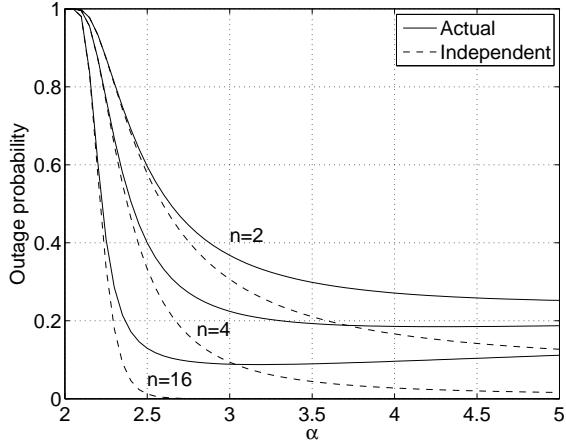,width=\figwidth}}
 \caption{Outage in SIMO system with selection combining and
 $2$, $4$, and $16$ receive antennas for $\theta=1$ and 
 $\Delta=\Gamma(1+\delta)\Gamma(1-\delta)/3$, where $\delta=2/\alpha$, as a function
 of the path loss exponent $\alpha$. The solid lines show the
 actual outage probability, while the dashed ones show the outage if there was no correlation
 in the interference.}
 \label{fig:simo}
 \end{figure}

In a selection combining scheme, a transmission is successful if
$\max_{k\in[n]} \{\sir_k\} >\theta$.
  The probability $p_n(\theta)$ that the SIR at at least one antenna exceeds the threshold 
 follows from \eqref{simo_joint} as
 \begin{equation}
  p_n(\theta)\triangleq\P\left(\bigcup_{k=1}^n S_k\right)=\sum_{k=1}^n (-1)^{k+1} 
  \binom nk P_k(\theta) \,. 
  \label{sel_combining}
 \end{equation}
Assuming independent interference, the probability of the same event would be
\[ \tilde p_n(\theta)=1-\left(1-e^{-\Delta\theta^\delta}\right)^n \,,\]
which differs substantially from \eqref{sel_combining}.
The gap between the outage probabilities $1-p_n(1)$ and $1-\tilde p_n(1)$ is illustrated in
 \figref{fig:simo}.
 While there is always a gain in increasing the number of antennas $n$,
 it is significantly smaller than under the assumption of independent interference.
 Also, it can be observed that the outage probability is no longer monotonically
 decreasing in $\alpha$ for all $n$.

\begin{figure}
 \centerline{\epsfig{file=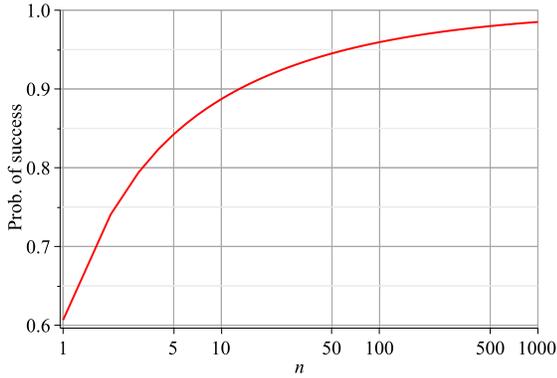,width=1.0\figwidth}}
 \caption{Success probability for selection combining as a function of the
 number of antennas $n$ on a logarithmic scale. $\Delta\theta^\delta=1/2$, $\delta=1/2$.}
 \label{fig:sel_comb}
 \end{figure}

While $\tilde p_n(\theta)\to 1$ quickly as $n\to\infty$, the asymptotic behavior of
$p_n(\theta)$ is less clear. A plot is shown in \figref{fig:sel_comb}.
We have the following result.
\begin{theorem}
For all $\Delta,\theta\geq 0$, $\delta\in(0,1)$,
\[ \lim_{n\to\infty} p_n(\theta)=1 \]
and, as $n\to\infty$,
\begin{equation}
  1-p_n(\theta) =\Omega(n^{-1-\epsilon})\,, \quad\forall\epsilon>0 \,. 
  \label{pn_cond}
\end{equation}
\end{theorem}
\begin{IEEEproof}
Conditioned on $\Phi$, the success probability goes to $1$ since
all events $S_k$ are independent (and have positive probability), \ie, $\lim_{n\to\infty} p_n(\theta\mid \Phi)=1$. Thus
\[ \E\big( \lim_{n\to\infty} p_n(\theta\mid \Phi)\big)=1\,, \]
which is the same as the desired limit $\lim_{n\to\infty} \E(p_n(\theta\mid \Phi))$
by monotone convergence.
For the bound on the tail probability, let $N(\theta)=\min\{k\colon \sir_k>\theta\}$
for $n=\infty$. We have
\begin{align}
\E N(\theta)&=\sum_{k=0}^\infty \P(N(\theta)>k)\nonumber \\
  &=\sum_{k=0}^\infty 1-p_k(\theta)\label{pk_limit} \,.
\end{align}
Replacing spatial diversity with temporal diversity, we can apply
\cite[Lemma 2]{net:Haenggi12tit} and set the transmit probability to $1$
(since in our case all interferers always transmit), and it follows that
$\E N(\theta)=\infty$.
So \eqref{pk_limit} diverges, which means that $1-p_k(\theta)$ decays
more slowly than $n^{-1-\epsilon}$ for any $\epsilon>0$.
\end{IEEEproof}

\subsection{The general two-antenna case}  
  \begin{corollary}
  The complete joint SIR distribution for $n=2$ is
\begin{align*}  
    \bar P_2(\theta_1,\theta_2)&\triangleq\P(\SIR_1<\theta_1, \,\SIR_2<\theta_2) \\
     &=1-\exp(-\Delta\theta_1^\delta)-\exp(-\Delta\theta_2^\delta)+\\
      &\qquad\exp\left(-\Delta \frac{\theta_1^{1+\delta}-\theta_2^{1+\delta}}{\theta_1-\theta_2}\right) \,.
 \end{align*}
\end{corollary}
\begin{IEEEproof}
 From Thm.~1, we obtain
  \begin{align}
    P_2(\theta_1,\theta_2)&\triangleq \P(\SIR_1>\theta_1,\SIR_2>\theta_2) \nonumber \\
  &=\exp\left(-\Delta \frac{\theta_1^{1+\delta}-\theta_2^{1+\delta}}{\theta_1-\theta_2}\right) \,
  \label{tilde_p2}
  \end{align}
  by the replacement
  \[ \left(\frac{1}{1+\theta_r\|x\|^{-\alpha}}\right)^2\;\rightarrow \; \left(\frac{1}{1+\theta_{r,1}\|x\|^{-\alpha}}\right)
  \left(\frac{1}{1+\theta_{r,2}\|x\|^{-\alpha}}\right) \]
  in the last two lines of the derivation in the proof. Since
  \begin{align*}  
    \bar P_2(\theta_1,\theta_2)&\triangleq\P(\SIR_1<\theta_1, \,\SIR_2<\theta_2) \\
    &=1-\P(\{\SIR_1>\theta_1\}\cup \{\SIR_2>\theta_2\}) \,,
 \end{align*}
the result follows from \eqref{tilde_p2}.
 \end{IEEEproof}
 For comparison,
  if interference was independent, the probability \eqref{tilde_p2} would be
  $\tilde P_2(\theta_1,\theta_2)=\exp(-\Delta(\theta_1^\delta+\theta_2^\delta))$.

\section{Conclusions}
We have derived the first results of the effect of interference correlation in Poisson 
networks with multi-antenna receivers. The diversity loss can be
quantified exactly using the {\em diversity polynomial}. 
Its effects are that
$\log P_n \propto n^\delta$
as opposed to
$\log \tilde P_n \propto n$ for independent interference, and that
the success probability in a selection combining scheme approaches $1$ at best
polynomially instead of exponentially.

The larger the path loss exponent (the smaller $\delta$), the more drastic
the effect of the interference correlation. Pearson's correlation coefficient
between the events that the SIR at two different antennas exceeds $\theta$
is approximately $1-\delta$ in the low-outage regime.

The results have important implications on the performance of multi-antenna
networks and raise interesting questions about how to best cope with
interference correlation.

 \bibliographystyle{IEEEtr}

 \end{document}